# Neutrino and graviton rest mass estimations by a phenomenological approach

Dimitar Valev

*Stara Zagora Department, Bulgarian Academy of Sciences, P.O. Box 73, 6000 Stara Zagora, Bulgaria*

**Abstract**

The ratio between the proton and electron masses is shown to be close to the ratio between the strong and electromagnetic interaction coupling constants at Extremely Low Energy (ELE). Based on the experimental data, this relation has been extended for the weak and gravitational interactions, too. Thus, a mass relation has been found, according to which the rest mass of the Lightest Free Massive Stable Particle (LFMSP), acted upon by a particular interaction, is proportional to the coupling constant of the respective interaction at ELE. On the basis of this mass relation, the electron neutrino and graviton masses have been approximately estimated to $2.1 \times 10^{-4}$ eV/c$^2$ and $2.3 \times 10^{-34}$ eV/c$^2$, respectively. The last value is of the order of the magnitude of $\frac{\hbar H}{c^2}$, where H is the Hubble constant. It is worth noting that this value has been obtained by fundamental constants only, without consideration of any cosmological models.

E-mail: valev@gbg.bg

## 1. Introduction

Although the neutrino and the graviton belong to different particle kinds (neutral lepton and carrier of the gravity, respectively), they have some similar properties. Both particles are not acted upon by the strong and the electromagnetic interactions, which makes their detection and investigation exceptionally difficult. Besides, both have masses that are many orders of magnitude lighter than the masses of the rest particles and they are generally accepted to be massless.

Decades after the experimental detection of the neutrino by Reines and Cowan [1], it was generally accepted that the neutrino mass $m_{0\nu}$ is rigorously zero. In the Fermi's theory of $\beta$-decay [2] as well as in the electroweak theory [3, 4] and hence, in the Standard Model (*SM*), the neutrinos have been accepted massless. Despite this, attempts to determine the neutrino mass have been made as early as it was found. The recent experiments bound $m_{0\nu}$ on the top and its upper limit has decreased millions of times in the latest experiments, as compared to the initial estimations of Pauli [5].

The first experiment, hinting that the neutrino probably possesses a mass, was carried out by Davis et al. [6]. The total flux of neutrinos from the Sun was found about three times lower than the one, predicted by theoretical solar models, thereby creating the problem for the solar neutrino deficit. This discrepancy can be explained if some of the electron neutrinos transform into another neutrino flavour. Within the frame of the *SM*, however, there is no place for massive neutrinos and neutrino oscillations. As a result, the detection of neutrino oscillations appears crucial for the *SM* and it requires its extension in direction to the *GUT*, *SUSY*, Superstring/M-theory and others.

Later, the experimental observations showed that the ratio between the atmospheric $\nu_\mu$ and $\nu_e$ fluxes is less than the theoretical predictions [7, 8]. This discrepancy became known as the atmospheric neutrino anomaly. Again it could be explained by the neutrino oscillations. The crucial experiments with the 50 *kton* neutrino detector Super-Kamiokande found strong evidence for oscillations (and hence - mass) in the atmospheric neutrinos [9].

The direct neutrino measurements allow to bound the neutrino mass. The upper limit for the mass of the lightest neutrino flavour $\nu_e$ was obtained from experiments for measurement of the high-energy part of the tritium $\beta$-spectrum and recent experiments yield ~ 2 $eV/c^2$ upper limit [10, 11]. As a result of the recent experiments, the upper mass limits of $\nu_\mu$ and $\nu_\tau$ reach 170 $keV/c^2$ [12] and 18.2 $MeV/c^2$ [13], respectively. The Solar and atmospheric neutrino experiments allow to find the neutrino mass squared differences $\Delta m_{12}^2 = m_2^2 - m_1^2$ and $\Delta m_{23}^2 = m_3^2 - m_2^2$, but not the absolute value of the neutrino masses. The astrophysical constraint of the neutrino mass is $\Sigma m_\nu$ < 2.2 $eV/c^2$ [14]. The recent extensions of *SM* lead to non-zero neutrino masses, which are within the large range from $10^{-6}$ $eV/c^2$ to 10 $eV/c^2$.

Although the graviton hasn't been experimentally detected yet, most of the quantum gravity models posit a neutral spin-2 particle, appearing carrier (gauge boson) of the gravity. Similarly to the case with the neutrino before 1998, the prevailing current opinion is that the graviton is massless. This opinion is connected with Einstein's theory of general relativity, where the gravity is described by a massless field of spin 2 in a generally covariance manner. The nonzero graviton mass produces a finite range of the gravity $r_g \sim \lambda_g = \frac{\hbar}{m_g c}$. There are two kinds of astrophysical methods for estimation of the upper limit of the graviton mass (or low limit of $\lambda_g$) – static and dynamic methods. The static methods are based on the search of difference between Yukawa potential for massive graviton and Newton potential for massless graviton. The Solar system measurements infer $\lambda_g > 2.8 \times 10^{15}$ *m*, that is equivalent to $m_g$ < $4.4 \times 10^{-22}$ $eV/c^2$ [15]. Rich galactic clusters allow to estimate $m_g$ < $2 \times 10^{-29}$ $h^{-1}$ $eV/c^2$ [16], where $h \approx 0.65$ is a dimensionless Hubble constant. This is the lowest limit of the graviton mass and therefore this value is used in the present paper. The dynamic methods are based on the differences of the emission and propagation of the gravitational waves from binary stellar systems in cases of massless or massive graviton. The possibilities of the astrophysical measurements to limit the graviton mass, including Laser Interferometer Space Antenna (*LISA*), are still of the order of the static tests magnitude in the Solar system [17]. Will and Yunes [18] suggest considerable improvement of these results for extra-Galactic massive binaries.

The theoretical estimations of the graviton mass are most often based on the assumption that the Compton wavelength of the graviton $\lambda_g$ is close to the Hubble distance $c/H \sim 1.4 \times 10^{26}$ *m*, which produces a value of the graviton mass $m_g \sim \frac{\hbar H}{c^2} \sim 10^{-33}$ $eV/c^2$, where $H \approx 65$ $km$ $s^{-1} Mpc^{-1}$ is the Hubble constant [19, 20]. Woodward et al. [21] obtain a value of the graviton mass $4.3 \times 10^{-34}$ $eV/c^2$ for an infinite stationary universe, but the expansion of the Universe is a fact, long ago established.

**2. Determination of the coupling constants at extremely low energy**

Four fundamental interactions are known - strong, electromagnetic, weak and gravitational, whose strength decreases in the direction from strong to gravitational interaction of dozens orders of magnitude. A measure for the interaction strength is a dimensionless quantity, namely the coupling constant of the interaction ($\alpha_i$), which is determined from the cross section of the respective processes. It is known that the coupling constants of the interactions are running [22, 23]. With increase of the processes energy, the coupling constant of the strong interaction decreases and reaches $\alpha_s \approx 0.11$ at 189 [24] *GeV*, and the rest coupling constants increase. Since the modern experiments are performed with energy of hundreds *GeV*, a value of the weak coupling constant, close on the electroweak scale, approaches the coupling constant of the electromagnetic interaction. The electromagnetic coupling constant increases exceptionally slow and remains $\alpha \approx 7.81 \times 10^{-3}$ even at energy $E \sim m_W c^2 \approx 80.4$ *GeV* [25]. The coupling constant would fulfill the role of unique property of the particular interaction when the energy is fixed. Recent unified theories predict unification of the four interactions on the Planck scale ($E \sim 10^{19}$ *GeV*). Close to such energy, the four interactions and coupling constants merge. Thus, with the energy increase, the interactions (and the coupling constants) converge and become more hardly differentiated from each other. A similar situation occurs with the particles since their rest masses become a negligible part of the full masses. The purpose of this paper is to relate the rest masses of the lightest stable particles, acted upon by the respective interactions with the respective coupling constants. That is why it is necessary to determine the coupling constants under conditions when the interactions (and the particles) are differentiated as much as possible and, this is the case when the energy of the processes reaches extremely low value. Thus, each coupling constant obtains a unique asymptotical value at Extremely Low Energy (*ELE*), which will be designated by $\alpha(0)$.

The coupling constant of the electromagnetic interaction $\alpha_e$ is known the fine structure constant $\alpha_e(0) = \frac{e^2}{\hbar c} \equiv \alpha \approx 7.30 \times 10^{-3}$, where *e* is the elementary electrical charge, $\hbar$ - Planck constant, *c* – the light velocity.

The coupling constant of the weak interaction $\alpha_w$ is determined by the expression:

$$\alpha_w = \frac{G_F m^2 c}{\hbar^3} \quad (1)$$

where $G_F$ is Fermi coupling constant, *m* – the interacting particle mass.

As mentioned above, it is necessary the coupling constants to be determined at *ELE* of the processes. The lowest-energy process, involving the weak interaction, is the neutron $\beta$ - decay. The process is running extremely slowly ($\tau \sim 880$ s), and a minimum quantity of energy is released $\Delta E \approx 0.78$ *MeV* $\sim m_e c^2 = 0.511$ *MeV*. Therefore, *m* in expression (1) is substituted by the electron mass $m_e$ and for the weak coupling constant at *ELE*, the following value is obtained:

$$\text{(2)} \quad \alpha_w(0) \sim \frac{G_F m_e^2 c}{\hbar^3} \approx 3.00 \times 10^{-12}$$

This value of the weak coupling constant is close to the one accepted in [26], where $\frac{\alpha_w}{\alpha} \sim 10^{-10}$. The value of the weak coupling constant, obtained from (2) is minimal as a result of the minimal energy of the neutron $\beta$ - decay. The rest decays involve the weak interaction, occurring with considerably higher energy and, as a result, the typical values of $\alpha_w$ are in orders of magnitude higher than such energy. For this reason $\alpha_w$ often determines by (1) replacing $m$ with the pion or proton mass.

It is known [27] that in cases of slow nucleons scattering (without angular momentum) the strong coupling constant reaches a maximum value $\alpha_s \sim 10 \div 17$, in [28] $\alpha_s$ is accepted $\sim 15$ and in [29] $\alpha_s \sim 14$. As a result, the strong coupling constant at *ELE* is accepted $\alpha_s(0) \sim 14 \pm 1$ in this paper. The strong coupling constant is often accepted a unit but this value corresponds to high energy of experiments within the *GeV* sector.

The coupling constant of the gravitational interaction is determined by the expression:

$$\text{(3)} \quad \alpha_g = \frac{G_N m_e m_p}{\hbar c} \approx 3.21 \times 10^{-42}$$

where $G_N$ is the universal gravitational constant and $m_e$ and $m_p$ are the electron and proton masses. This estimation is a conventional medial value between $\alpha_g = \frac{G_N m_e^2}{\hbar c}$ and $\alpha_g = \frac{G_N m_p^2}{\hbar c}$.

Thus the coupling constant of the strong, electromagnetic, weak and gravitational interactions at *ELE* are 14, $7.30 \times 10^{-3}$, $3.0 \times 10^{-12}$ and $3.21 \times 10^{-42}$, respectively.

**3. Phenomenological mass relation for free massive stable particles**

Among the hundreds observed particles, only several free particles are notable, which are stable or at least their lifetime is longer than the age of the Universe. These particles are the proton (*p*), electron (*e*), photon (*γ*), neutrinos ($v_i$) and hypothetical graviton (*g*). These particles play a key role since together with the quasi-stable neutron they build the known matter in the Universe. Although according to the quantum chromodynamics the proton is composed by quarks, it has displayed itself as an undivided particle in the recent experiments. Only *free massive stable* particles are examined in this paper. The quarks and gluons are bound in the hadrons by confinement and they cannot be immediately detected in the experiments, and the photon is massless. Therefore, these particles are not the subject of this paper.

The ratio between the proton and electron masses is $\frac{m_p}{m_e} \approx 1836$. On the other side, the ratio between the strong and electromagnetic coupling constants at *ELE* is $\frac{\alpha_s(0)}{\alpha} \approx 1918$. The two ratios differentiate by less than 5%, which is less than the error of $\alpha_s(0)$. Consequentially, the found relation is hardly an incidental coincidence, therefore would be written:

$$(4) \qquad \frac{m_p}{m_e} \approx \frac{\alpha_s(0)}{\alpha}$$

Thus, it is shown that the ratio between the proton and electron masses is close to the ratio between the strong and electromagnetic coupling constants at ELE. The proton and electron are the Lightest Free Massive Stable Particles (*LFMSP*), acted upon by the strong and electromagnetic interactions, respectively. This relation is important since it connects the masses of *LFMSP*, acted upon by the strong and electromagnetic interactions and the respective coupling constant at *ELE*. The relation (4) suggests that the mass of *LFMSP*, acted upon by the strong (or the electromagnetic) interaction is proportional to the respective coupling constant at *ELE*, i.e. $m_p \approx k\alpha_s(0)$ and $m_e \approx k\alpha$, where *k* is a constant. Therefore, it is interesting to examine whether this rule is valid both for the weak interaction, which is several orders of magnitude weaker than the electromagnetic interaction and for the gravity, which is dozens orders of magnitude weaker than the weak interaction. *LFMSP* acted upon by the weak interaction is the electron neutrino and *LFMSP* acted upon by the gravity most probably appears the hypothetical graviton. Although the rest masses of the two particles are yet unknown, the direct neutrino mass experiments and the theoretical models suggest the $\nu_e$ mass between $10^{-6}$ *eV/c²* and 2 *eV/c²*, i.e. $\nu_e$ is several orders of magnitude lighter than the electron. Again, the astrophysical constraints allow to find the upper limits of the graviton mass and according to these constraints, if the graviton really exists its mass would be less than $3.1 \times 10^{-29}$ *eV/c²*, i.e. dozens orders of magnitude lighter than $\nu_e$. Table I presents the obtained in Section 2 coupling constants of the interactions at *ELE*, as well as the masses of *LFMSP*, acted upon by the respective interactions. The experimental upper limits of the electron neutrino and graviton masses are presented, too.

Table I shows that with increasing the interaction strength (coupling constant), the mass $m_{min}$ of *LFMSP* acted upon by the respective interaction also increases. The data from Table I are presented in a double-logarithmic scale in Fig.1, which shows that the trend is clearly expressed.

TABLE I: Coupling constants of interactions at *ELE* and the masses of *LFMSP* acted upon by the respective interaction.

| Fundamental Interaction | Coupling Constant | Particle | Experimental Mass ($eV/c^2$) | Calculated Mass ($eV/c^2$) |
|---|---|---|---|---|
| Strong | 14 | $p$ | $9.38 \times 10^8$ | $9.8 \times 10^8$ |
| Electromagnetic | $7.30 \times 10^{-3}$ | $e$ | $5.11 \times 10^5$ | - |
| Weak | $3.00 \times 10^{-12}$ | $\nu_e$ | $0 < m < 2$ | $2.1 \times 10^{-4}$ |
| Gravitational | $3.21 \times 10^{-42}$ | $g$ | $< 3.1 \times 10^{-29}$ | $2.3 \times 10^{-34}$ |

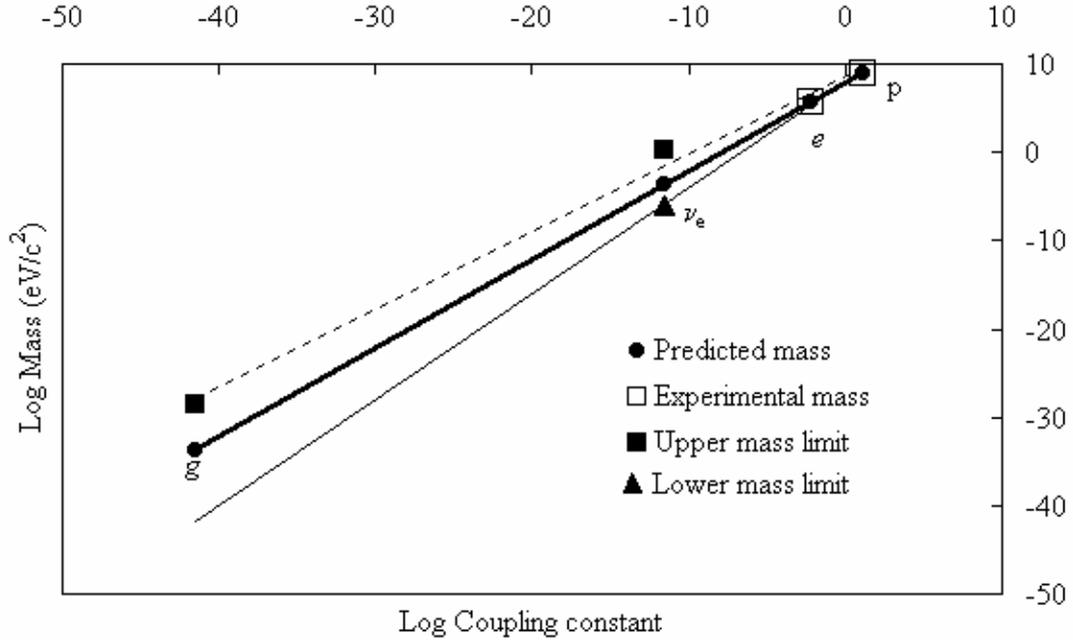

Fig. 1: Dependence of the mass of *LFMSP* acted upon by a particular interaction on the coupling constant of the respective interaction at *ELE*. The dashed line presents the approximation (5) of $e$, $p$ and the upper mass limits of $g$ and $\nu_e$. The thin solid line presents the approximation (6) of $e$, $p$ and the lowest limit mass of $\nu_e$ taken from the models. The thick solid line presents the strict linear approximation (S=1).

The points in Fig. 1, corresponding to the electron and proton masses and to the upper limit masses of the electron neutrino and graviton, are approximated by the least squares with a power law:

$$(5) \qquad Log\, m_{min} = 0.880\, Log\, \alpha(0) + 8.493$$

Although this approximation is only on 4 points, the correlation found is close and the correlation coefficient reaches $r = 0.997$, which supports the power law. The slope ($S$) is a little smaller than a unit, but it should be reminded that instead of the electron neutrino and graviton masses, their upper limit values are used, which produce a certain underestimation of the $S$ value. Therefore it can be said that the regression is close to a linear one. Analogously, the points, corresponding to the electron and proton masses and to the low limit of the electron neutrino mass from the theoretical models are approximated with a power law, too:

$$(6) \qquad Log\, m_{min} = 1.196\, Log\, \alpha(0) + 7.883$$

This approximation is similar to (5) and the correlation is close again ($r = 0.999$). Now the slope is a little bigger than a unit. The certain slope overestimation is due to the fact that in this case instead of the still unknown value of the electron neutrino mass, the lowest neutrino mass limit is used from the models. Approximations (5) and (6) produce at $\alpha_g(0) = 3.21 \times 10^{-42}$ a value for the graviton mass in an interval from $1.8 \times 10^{-42}$ $eV/c^2$ to $3.1 \times 10^{-29}$ $eV/c^2$, i.e. it suggests that the graviton possesses a negligible, yet a *nonzero mass*. These approximations show that the mass of *LFMSP*, acted upon by a particular interaction, increases with the increase of the respective coupling constant $\alpha(0)$ by a power law with $S \sim 1$, i.e. close to the linear one. The coincidence (4) of the ratio between the proton and electron masses with the ratio between the strong and electromagnetic coupling constants at *ELE* also supports a linear dependence (without intercept) between $m_{min}$ and $\alpha(0)$. Hence, the experimental data suggest a linear dependence ($S = 1$) between the mass of *LFMSP* and the coupling constants:

$$(7) \qquad Log\, m_{min} = Log\, \alpha(0) + k_0$$

where $k_0$ is a constant.

Expression (7) transforms into $m_{min} = 10^{k_0} \alpha(0) = k\alpha(0)$. In this way the experimental data and constraints suggest that the mass of *LFMSP*, acted upon by a particular interaction, is proportional to the coupling constant of the respective interaction at *ELE*:

$$m_{i\min} = k\alpha_i(0) \qquad (8)$$

where $k$ is a constant, $i = 1, 2, 3, 4$ - index for each interaction and *LFMSP* acted upon by a respective interaction.

Constant $k$ can be determined by the fine structure constant ($\alpha_1 \equiv \alpha$) and the electron mass ($m_{1min} \equiv m_e$) since both are measured with very high precision $k = \frac{m_e}{\alpha} \approx 7.00 \times 10^7 \, eV/c^2$. The substitution of this value in (8) yields the mass relation:

$$m_{i\min} = \frac{m_e}{\alpha}\alpha_i(0) \qquad (9)$$

Therefore, *LFMSP* corresponds to each fundamental interaction and its mass is proportional to the strength of the respective interaction.

### 4. Neutrino and graviton mass estimations and discussions

The found mass relation (9) can be examined by the strong interaction since the proton mass is measured with high precision. The application of the mass relation on the strong interaction yields the lightest stable hadron mass $m_p \approx 9.80 \times 10^8 \, eV/c^2$. Therefore, the proton mass value obtained by the mass relation would be only 4.5% higher than the experimental value of $m_p$. This result shows that the found mass relation possesses heuristic power. The application of the mass relation (9) on the weak interaction allows to evaluate the mass of the electron neutrino $m_{ve} \approx 2.1 \times 10^{-4} \, eV/c^2$. This value is close to the prediction of the simple *SO(10)* model for the lightest neutrino mass $m_{ve} = 2.4 \times 10^{-4} \, eV/c^2$ [30].

Finally, the application of the mass relation (9) on the gravitational interaction produces an estimation of the graviton mass $m_g = \frac{m_e}{\alpha}\alpha_g = \frac{G_N m_e^2 m_p}{e^2} \approx 2.3 \times 10^{-34} \, eV/c^2$, which is several orders of magnitude smaller than the upper limit of the graviton mass, obtained by astrophysical constraints. Thus, the graviton mass has been estimated by *fundamental constants* only - universal gravitational constant, electron and proton masses and elementary electrical charge. It is worth noting that this value is of the order of the magnitude of $\frac{\hbar H}{c^2}$, but by suggested approach has been obtained without consideration of any cosmological models.

The calculated masses of the four free stable particles are presented in the last column of Table I. It shows that the fitting of the calculated values and the experimental data is satisfactory. In this way the found mass relation is remarkable since it unifies the masses of four stable particles of completely different kinds (*p*, *e*, $\nu_e$ and *g*) and covers an extremely wide range of values, exceeding 40 orders of magnitude. The found mass relation allows approximate estimation of the neutrino and graviton masses, which affords the opportunity for its verification in the nearest years when the direct neutrino mass measurements are expected to reach the necessary threshold of sensitivity.

The obtained value $m_{\nu e} \approx 2.1 \times 10^{-4}$ $eV/c^2$ and the results from the solar and atmospheric neutrino experiments allow to estimate the masses of the heavier neutrino flavours - $\nu_\mu$ and $\nu_\tau$. The results from the Super Kamiokande experiment lead to the neutrino mass squared difference $\Delta m_{23}^2 \sim 2.7 \times 10^{-3} eV^2$ [31]. Recent results on solar neutrinos provide hints that the Large Mixing Angle (*LMA*) Mikheyev-Smirnov-Wolfenstein (*MSW*) solution is more probable than Small Mixing Angle (*SMA*) *MSW* [32]. The *LMA* leads to $\Delta m_{12}^2 \sim 7 \times 10^{-5} eV^2$ [33] and the *SMA* leads to $\Delta m_{12}^2 \sim 6 \times 10^{-6} eV^2$ [34]. In this way both solutions yield $m_{\nu \tau} \approx 0.05$ $eV/c^2$. The most appropriate *LMA* yields $m_{\nu \mu} \approx 8.4 \times 10^{-3}$ $eV/c^2$, and *SMA* yields $m_{\nu \mu} \approx 2.5 \times 10^{-3}$ $eV/c^2$. Thus, the obtained values of the neutrino masses support the normal hierarchy of neutrino masses.

Now it is not clear yet what the cause for the relationship between the masses of *LFMSP* and the interaction coupling constants is but its existence is confirmed by the recent experimental data and constraints. Most probably the found mass relation represents an expression of a universal symmetry, including free stable particles of most diverse kinds (hadron, charged lepton, neutral lepton and carrier of the gravity).

It is interesting that the mass relation (9) is very similar to equation (10) derived in [35] by a totally different approach, namely by strong gravity model and astrophysical constraints:

$$(10) \qquad m = \left( \frac{\hbar^2}{G_N R_0} \right)^{1/3} \frac{q^2}{\hbar c}$$

where $R_0 = c/H$ is the Hubble distance and *q* is the "main charge".

The mass dimension coefficient of proportionality $m_0 = \left( \frac{\hbar^2}{G_N R_0} \right)^{1/3} \sim 1.05 \times 10^{-28}$ *kg* is close to $k = \frac{m_e}{\alpha} \approx 1.25 \times 10^{-28}$ *kg*.

The presence of an exceptionally small, yet nonzero mass of the graviton, involves a finite range of the gravity $r_g \sim \lambdabar_g$ and Yukawa potential of the gravitational field $\phi(r) = -\dfrac{G_N M}{r}\exp(-r/\lambdabar_g)$, where $\lambdabar_g$ is Compton wavelength of the graviton $\lambdabar_g = \dfrac{\hbar}{m_g c} \sim c/H$. Since $\lambdabar_g$ is in order of magnitude of the Hubble distance, the deviation of Newton potential from Yukawa potential is manifested very weakly at a distance $r \ll c/H$. As a result, the experimental determination of the graviton mass will be a serious challenge. Yet, it can be expected that appropriate astrophysical or laboratory experiments will be found for experimental determination of the graviton mass. The massive graviton might turn of considerable importance for the description of the processes in the nuclei of the active galaxies and quasars, the gravitational collapse as well as for the improvement of the cosmological models.

The massive graviton places other challenges before the modern unified theories. Among them are the famous van Dam-Veltman-Zakharov (*vDVZ*) discontinuity [36, 37] and the violation of the gauge invariance and the general covariance. Yet, there are already encouraging attempts to solve *vDVZ* discontinuity in anti de Sitter background [38].

## 5. Conclusions

Among the huge multitude of observed particles, a small group of several free particles is noticeable, which are stables or, at least their lifetimes are longer than the age of the Universe. These particles are very important since together with the quasi-stable neutron they build the known matter in the Universe.

The interaction coupling constants are determined at *ELE* close to $m_e c^2$ when the interactions are most different in their strength and are differentiated clearly from one another. It is shown that the ratio between the proton and electron masses is close to the ratio between the strong and electromagnetic coupling constants at *ELE*. Further on, based on the experimental data, a mass relation has been found, according to which the rest mass of the Lightest Free Massive Stable Particle (LFMSP), acted upon by a particular interaction, is proportional to the coupling constant of the respective interaction at ELE. This mass relation is remarkable, since it connects the masses of the particles, most important for the stability of the Universe, i.e. the stable particles, with the most substantial property of the fundamental interactions, i.e. their coupling constants. The applied approach suggests that the graviton mass is nonzero. The electron neutrino and graviton masses are approximately estimated to $2.1 \times 10^{-4}$ *eV/c²* and $2.3 \times 10^{-34}$ *eV/c²*, respectively. The obtained value of the graviton mass is of the order of the magnitude of $\dfrac{\hbar H}{c^2}$ and this value has been found by fundamental constants only, without consideration of any cosmological models. The masses of the heavier neutrinos $v_\mu$ and $v_\tau$ are estimated by the results of the solar and atmospheric neutrino experiments. The mass of $v_\tau$ is close to 0.05 $eV/c^2$ while the estimation of $m_{v_\mu}$ depends to a certain extent on the *MSW* solution. Yet, both solutions support the normal hierarchy of the neutrino masses.

**Acknowledgements**

I would like to thank Prof. I. G. Koprinkov for his useful discussions.